\documentclass[twoside,twocolumn,9pt]{article}
\usepackage{extsizes}
\usepackage[super,sort&compress,comma]{natbib} 
\usepackage[version=3]{mhchem}
\usepackage[left=1.5cm, right=1.5cm, top=1.785cm, bottom=2.0cm]{geometry}
\usepackage{balance}
\usepackage{mathptmx}
\usepackage{sectsty}
\usepackage{graphicx} 
\usepackage{lastpage}
\usepackage[format=plain,justification=justified,singlelinecheck=false,font={stretch=1.125,small,sf},labelfont=bf,labelsep=space]{caption}
\usepackage{float}
\usepackage{fancyhdr}
\usepackage{fnpos}
\usepackage[english]{babel}
\addto{\captionsenglish}{%
  
}
\usepackage{array}
\usepackage{droidsans}
\usepackage{charter}
\usepackage[T1]{fontenc}
\usepackage[usenames,dvipsnames]{xcolor}
\usepackage{setspace}
\usepackage[compact]{titlesec}
\usepackage{hyperref}
\usepackage{ulem}
\usepackage{epstopdf}%This line makes .eps figures into .pdf - please comment out if not required.

\definecolor{cream}{RGB}{222,217,201}

\begin{document}

\pagestyle{fancy}
\thispagestyle{plain}
\fancypagestyle{plain}{
%%%HEADER%%%
\renewcommand{\headrulewidth}{0pt}
}
%%%END OF HEADER%%%

%%%PAGE SETUP - Please do not change any commands within this section%%%
\makeFNbottom
\makeatletter
\renewcommand\LARGE{\@setfontsize\LARGE{15pt}{17}}
\renewcommand\Large{\@setfontsize\Large{12pt}{14}}
\renewcommand\large{\@setfontsize\large{10pt}{12}}
\renewcommand\footnotesize{\@setfontsize\footnotesize{7pt}{10}}
\makeatother

\renewcommand{\thefootnote}{\fnsymbol{footnote}}
\renewcommand\footnoterule{\vspace*{1pt}% 
\color{cream}\hrule width 3.5in height 0.4pt \color{black}\vspace*{5pt}} 
\setcounter{secnumdepth}{5}

\makeatletter 
\renewcommand\@biblabel[1]{#1}            
\renewcommand\@makefntext[1]% 
{\noindent\makebox[0pt][r]{\@thefnmark\,}#1}
\makeatother 
\renewcommand{\figurename}{\small{Fig.}~}
\sectionfont{\sffamily\Large}
\subsectionfont{\normalsize}
\subsubsectionfont{\bf}
\setstretch{1.125} %In particular, please do not alter this line.
\setlength{\skip\footins}{0.8cm}
\setlength{\footnotesep}{0.25cm}
\setlength{\jot}{10pt}
\titlespacing*{\section}{0pt}{4pt}{4pt}
\titlespacing*{\subsection}{0pt}{15pt}{1pt}
%%%END OF PAGE SETUP%%%

%%%FOOTER%%%
\fancyfoot{}
\fancyfoot[LO,RE]{\vspace{-7.1pt}\includegraphics[height=9pt]{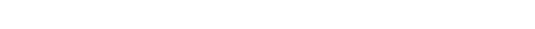}}
\fancyfoot[CO]{\vspace{-7.1pt}\hspace{11.9cm}\includegraphics{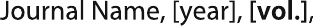}}
\fancyfoot[CE]{\vspace{-7.2pt}\hspace{-13.2cm}\includegraphics{head_foot/RF}}
\fancyfoot[RO]{\footnotesize{\sffamily{1--\pageref{LastPage} ~\textbar  \hspace{2pt}\thepage}}}
\fancyfoot[LE]{\footnotesize{\sffamily{\thepage~\textbar\hspace{4.65cm} 1--\pageref{LastPage}}}}
\fancyhead{}
\renewcommand{\headrulewidth}{0pt} 
\renewcommand{\footrulewidth}{0pt}
\setlength{\arrayrulewidth}{1pt}
\setlength{\columnsep}{6.5mm}
\setlength\bibsep{1pt}
%%%END OF FOOTER%%%

\newcommand{\bl}[1]{\color{blue}{#1}}
\newcommand{\red}[1]{\color{red}{#1}}
\newcommand{\gr}[1]{\color{green}{#1}}

%%%FIGURE SETUP - please do not change any commands within this section%%%
\makeatletter 
\newlength{\figrulesep} 
\setlength{\figrulesep}{0.5\textfloatsep} 

\newcommand{\topfigrule}{\vspace*{-1pt}% 
\noindent{\color{cream}\rule[-\figrulesep]{\columnwidth}{1.5pt}} }

\newcommand{\botfigrule}{\vspace*{-2pt}% 
\noindent{\color{cream}\rule[\figrulesep]{\columnwidth}{1.5pt}} }

\newcommand{\dblfigrule}{\vspace*{-1pt}% 
\noindent{\color{cream}\rule[-\figrulesep]{\textwidth}{1.5pt}} }

\makeatother
%%%END OF FIGURE SETUP%%%

%%%TITLE, AUTHORS AND ABSTRACT%%%
\twocolumn[
  \begin{@twocolumnfalse}
{\includegraphics[height=30pt]{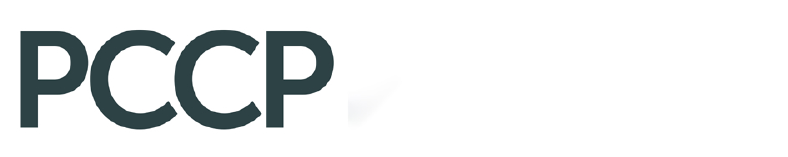}\hfill\raisebox{0pt}[0pt][0pt]{\includegraphics[height=55pt]{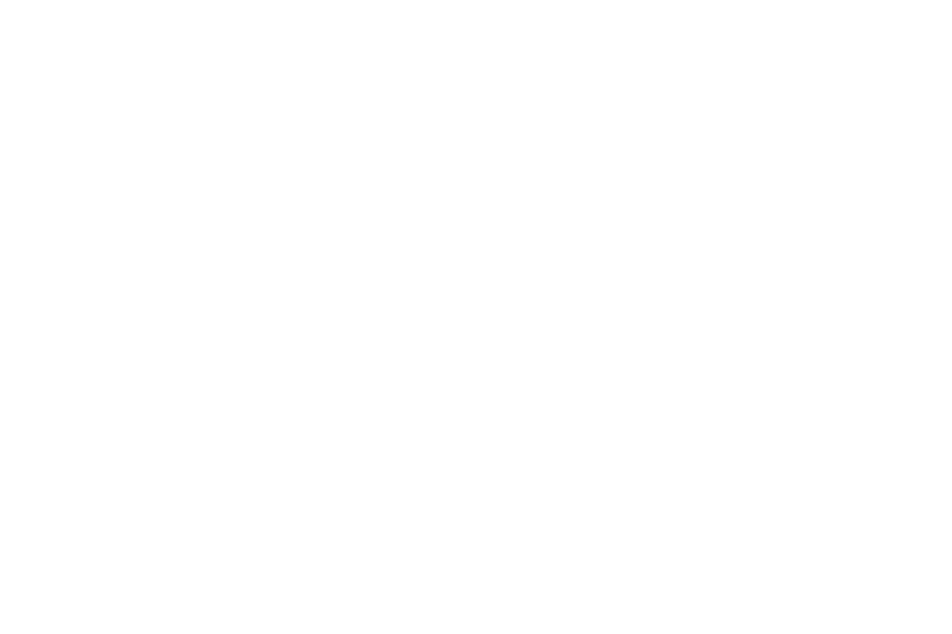}}\\[1ex]
\includegraphics[width=18.5cm]{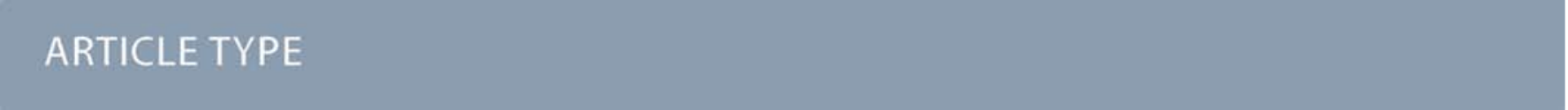}}\par
\vspace{1em}
\sffamily
\begin{tabular}{m{4.5cm} p{13.5cm} }

\includegraphics{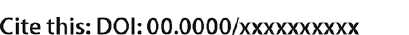} & \noindent\LARGE{\textbf{Exploring a novel class of Janus MXenes by first principles calculations: structural, electronic and magnetic properties of Sc$_2$CXT, X=O, F, OH; T=C, S, N$^\dag$}} \\%Article title goes here instead of the text "This is the title"
\vspace{0.3cm} & \vspace{0.3cm} \\

 & \noindent\large{S.~\"Ozcan$^{\ast}$\textit{$^{a}$} and B.Biel{$^{b\ddag}$}}\\%Author names go here instead of "Full name", etc.

\includegraphics{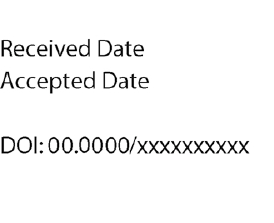} & \noindent\normalsize{The already intriguing electronic and optical properties of the MXene Sc$_2$C family can be further tuned through a wide range of possible functionalizations. Here, by means of Density Functional Theory, we show that the 36 possible elements of the Janus MXT (M:Sc$_2$C, X:O, F, OH, T:C, N, S) family, built by considering the four possible structural models (i) FCC, (ii) HCP , (iii) FCC + HCP, and (iv) HCP + FCC, are all potentially stable. The analysis of their mechanical properties shows the excellent mechanical flexibility of functionalized MXenes (f-MXenes) under large strain, making them more suitable for applications where stress could be an issue. Interestingly, while Sc$_2$C presents a metallic character, Sc$_2$COS, Sc$_2$CFN and Sc$_2$COHN are found to be semiconductors with bandgaps of 2.5 eV (indirect), 1.67 eV (indirect) and 1.1 eV (direct), respectively, which presents promising applications for nano- and optoelectronics. Moreover, Sc$_2$CFC presents a ferromagnetic ground state with the 2x2x1 supercell magnetic moment of 3.99 $\mu$B, while the ground state of Sc$_2$COHC might be antiferromagnetic with a magnetic moment of 3.98 $\mu$B, depending on the environment. Remarkably, the band structures of Sc$_2$CFC and Sc$_2$COHC present a half-metallic character with an HSE06 fundamental band gap of 0.60 eV and 0.48 eV, respectively. Our results confirm the extraordinary potential of the Janus MXT (M:Sc$_2$C, X:O, F, OH, T:C, N, S) family for novel applications in 2D nano-,opto- and spintronics.} \\

\end{tabular}

 \end{@twocolumnfalse} \vspace{0.6cm}

  ]
%%%END OF TITLE, AUTHORS AND ABSTRACT%%%

%%%FONT SETUP - please do not change any commands within this section
\renewcommand*\rmdefault{bch}\normalfont\upshape
\rmfamily
\section*{}
\vspace{-1cm}

%%%FOOTNOTES%%%

\footnotetext{\textit{$^{a}$~Department of Physics, Aksaray University, 68100 Aksaray, Turkey. E-mail: sozkaya@aksaray.edu.tr}}
\footnotetext{\textit{$^{b}$~ Department of Atomic, Molecular and Nuclear Physics \& Instituto Carlos I de F\'isica Te\'orica y Computacional, Faculty of Science, Campus de Fuente Nueva, University of Granada, 18071 Granada, Spain. }}

%Please use \dag to cite the ESI in the main text of the article.
%If you article does not have ESI please remove the the \dag symbol from the title and the footnotetext below.
\footnotetext{\dag~Electronic Supplementary Information (ESI) available: [details of any supplementary information available should be included here]. See DOI: 10.1039/cXCP00000x/}
%additional addresses can be cited as above using the lower-case letters, c, d, e... If all authors are from the same address, no letter is required

\footnotetext{\ddag~Additional footnotes to the title and authors can be included \textit{e.g.}\ `Present address:' or `These authors contributed equally to this work' as above using the symbols: \ddag, \textsection, and \P. Please place the appropriate symbol next to the author's name and include a \texttt{\textbackslash footnotetext} entry in the the correct place in the list.}

%%%END OF FOOTNOTES%%%

%%%MAIN TEXT%%%%

\section{Introduction}
The unique properties triggered by the reduced 
dimensionality~\cite{Chen-2015,Jin-2016,Mak-2010,Jing-2013} of two-dimensional (2D) electronic materials have been object of exhaustive investigations since 
the discovery of graphene by Novoselov et al.~\cite{Novoselov-2004}. 
Several families of 2D materials, such as transition metal 
dichalcogenides (TMDs)~\cite{Coleman-2011}, hexagonal boron 
nitrides~\cite{Novoselov-2005}, few-layer metal oxides
~\cite{Sreedhara-2013}, or metal chalcogenides (e.g. MoS$_2$~\cite{Joensen-1986}, 
WS$_2$~\cite{Seo-2007}) have been recently identified and investigated 
to be used in semiconductor devices, rechargeable ion batteries, 
and catalysis devices. Recently, a new family of 2D transition metal 
carbides and carbonitrides (MXenes) have received special attention 
due to their potential applications in sensors, electronic devices, 
catalysis, storage systems, energy conversion, and spintronics 
~\cite{Anasori-2015,Gao-2016,Zhang2017,Pandey-2017,Zha-2016,Liang-2017,Fu-2017,Akgenc2021,Abdullahi2022,Siriwardane2021}. 
MXenes are generally synthesized by selectively etching 
“A” layers from M$_{n+1}$AX$_n$ phases  
~\cite{Naguib-2012,Naguib2013,Lukatskaya-2013,Mashtalir-2013,Tang-2016}, 
where M is an early transition metal, A is an element from the  
IIIA or IVA groups, and X is either carbon or nitrogen (n =1, 2, or 3). 
“A” atoms are removed from the MAX phase during the etching process.

For MXenes, surface termination with functional groups (-O, -OH, or -F) passivate 
the outer-layer metal atoms. Many theoretical studies show that most of 
the MXenes are metallic and that the majority of them still remain metallic after 
surface passivation. Exceptionally, while pristine monolayer Sc$_2$C is 
metallic, monolayer Sc$_2$CT$_2$ (T = OH, F, and O) are semiconductors with 
a bandgap of 0.45–-1.80 eV after surface functionalization (-O, -OH, or -F) 
~\cite{Yorulmaz-2016,Wang-2017}, which opens an exciting path towards the controlled design of 2D MXenes with selected properties. Besides the impact on the electronic structure, surface functionalization might
also effect other properties such as the thermodynamic stability, elasticity, and electronic and thermal transport 
of the MXenes ~\cite{Zhang-2017,Zha-2016,Eames-2014}. For instance, S-terminated 
MXenes have lower diffusion barriers (related to the electronic mobility and charge/discharge rate of the electrode material)
than their O-functionalized counterparts ~\cite{Zhu-2016,Li-2020}, hence providing new candidates for the selection of the most adequate electrode material for batteries. Shao {\it et al.} investigated the stability and electronic properties of nitrogen-functionalized MXenes with the help of a detailed 
\textsl{ab initio} pseudopotential calculation ~\cite{Shao-2017}. Their results indicate that Nb$_2$CN$_2$ 
and Ta$_2$CN$_2$ Mxenes are direct bandgap semiconductors with ultrahigh carrier mobility. Following this result, Xiao {\it et al.} \cite{Xiao-2019} presented first-principles total-energy 
calculations that proved that Ti$_2$C and its derivatives,Ti$_2$CC$_2$, Ti$_2$CO$_2$, 
and Ti$_2$CS$_2$ have excellent potential for their use as anodes in Na-ion batteries.

Following these enticing results, Janus materials created from MXenes with different surface termination (i. e., different functional groups on each of the two surfaces) have already been investigated for varied applications
\cite{Jin-2018,Wang-2019,Akgen-2020,Xiong-2020}. Due to the two different 
chemical surfaces, the mirror asymmetry of Janus materials introduces an out-of-plane structural symmetry, which can lead to new properties such as an intrinsic out-of-plane electric field or spin-orbit-induced spin splitting. These exciting possibilities have generated a surge of interest in these materials \cite{Khazaei-2017}.

Structural symmetry-breaking, as the one found in 2D Janus MXenes, which present different terminations on each of their two surfaces, is key in defining the electronic properties of two-dimensional materials. Many efforts have been taken, for instance, to breaking the in-plane symmetry of graphene (see, for example, Zhang {\it et al.} ~\cite{Zhang-09} and references therein), and van der Waals heterostructures (see, for instance, Hunt {\it et al.} ~\cite{Hunt-13}). The direct electronic band gaps of transition metal dichalcogenide monolayers are due indeed to this in-plane inversion asymmetry. Additionally, it has been shown that, by breaking the out-of-plane mirror symmetry with either external electric fields ~\cite{Yuan-13, Wu-13}, or with an asymmetric out-of-plane structural configuration ~\cite{Cheng-13}, similar to the one that exists in 2D Janus MXenes, an additional degree of freedom allowing spin manipulation can be obtained. Hence, creating such asymmetry can lead to a wealth of physical phenomena not present in standard MXenes. For instance, while silicene lacks a sizeable energy gap, thus preventing it from being used as a semiconductor in 2D nanodevices, Sun {\it et al}.  reported that the Janus silicene, a silicene monolayer asymmetrically functionalized with hydrogen on one side and halogen atoms on the other, exhibits good kinetic stability, with a band gap that can be tuned between 1.91 and 2.66 eV ~\cite{Sun-16}. 
 %Other study showed that the asymmetrically functionalized Ti$_2$C with -OCH$_3$ and -OH groups, Ti$_2$C(OCH$_3$)(OH), exhibits a high stability  ~\cite{Enyashin-13}.
 Based on Density Functional Theory (DFT) calculations, it has been reported that the band gap of other asymmetrically functionalized MXenes such as Janus Cr$_2$C – Cr$_2$CXX' (X, X' = H, F, Cl, Br, OH) can be effectively tuned by a selection of a suitable pair of chemical elements (functional groups) ending the upper and the lower surfaces ~\cite{He2016}. State-of-the-art DFT calculations found that the monolayer Janus transition metal dichalcogenides (JTMDs) WSSe and WSeTe exhibit superior mobilities than conventional TMD monolayers such as MoS$_2$ ~\cite{Wang2018}.

Some of these Janus 2D materials have already been synthesized. For instance, in 2013, Janus graphene was successfully fabricated by Zhang {\it et al.} \cite{Zhang-2013}. In subsequent works,Janus-functionalized graphene (asymmetrically terminated by H, F, O) was synthesized by several groups \cite{Karlicky-2013,Holm-2018}. Recently, Zhang {\it et al.} ~\cite{Zhang-2017} developed a Janus MoSSe by substituting the top Se layer of MoSe$_2$ with S atoms. This approach, which breaks the out-of-plane structural symmetry of transition metal dichalcogenide monolayers, was also followed by Lu {\it et al}\cite{Lu17}, who found that this material presents an optically active vertical dipole, making it a 2D platform to study light–matter interactions where dipole orientation is critical, such as dipole–dipole correlations and strong coupling with plasmonic structures. %\sout{reported a synthetic strategy to grow the Janus structure of MoSSe directly by means of scanning transmission electron microscopy and energy-dependent X-ray photoelectron spectroscopy}~\cite{Lu-2017}. 
Recently, thermodynamically stable 2D Janus monolayers SWSe, SNbSe, and SMoSe were synthesized for the first time by Qin {\it et al.}~\cite{Qin-22}, showing excellent excitonic and structural quality and making them very valuable for optoelectronic applications.

Inspired by the experimental achievements of Janus 2D materials, the synthesis of Janus MXenes with asymmetrically surface-functionalized groups is expected in the near future.

As the functionalization of quality surfaces plays an essential role in many applications, such as high-speed electronic devices ~\cite{Pang-19}, the search for new functional groups or atoms to enable the design of Janus MXenes for next-generation devices becomes mandatory. Besides, experimentally obtained MXenes are usually covered with functional groups such as O, -OH, and F. Indeed, sulfur is a promising functional group for MXenes owing to its earth abundance, low cost, and environmental benignity, and S-terminated MXenes could be thus employed as anode material in Na- and K-based batteries. The aim of this work is hence to provide a comprehensive study of all possible 2D Janus materials based on the Sc$_2$CX MXene (X=O, F, OH) when functionalized with some of its most common functional groups (T=C, S, N) on both surfaces. We must note, however, that structures with mixed passivation might also be obtained experimentally. Further work is necessary to assess their relative stability with respect to the structures studied here and to determine their electronic, mechanical and magnetic properties. Our results will help to identify those structurally stable materials with the desired electronic properties based on their electronic character, gap size, and magnetic properties.
\section{Method}
All the calculations were carried out with the use of the {\it Vienna  Ab initio}
Simulation Package (VASP)~\cite{vasp1,htt,Kre-99,Kres-96}, based on
Density Functional Theory (DFT). In this method, the Kohn--Sham single particle functions are
expanded based on plane waves up to the cut-off energy of 51 Ry. The lattice constants were 
optimized and atoms were relaxed without any constraint until the Hellmann-Feynman forces became less than 0.003 eV/ {\AA}. 
The Fermi level Gaussian smearing factor was set to 0.05 eV.
The electron--ion interactions were described by using the projector
augmented-wave (PAW) method~\cite{Kre-99,Bl-94}. For electron
exchange and correlation terms, a Perdew--Zunger-type
functional~\cite{Per-81,Per-92} was used within the generalized
gradient approximation (GGA)~\cite{Bl-94} in its PBE parameterization 
~\cite{Perdew-1996}. To overcome the well-known underestimation of the band gap values by this functional, we have refined our calculations by using the Heyd–Scuseria–Ernzerhof (HSE06) functional, which has been proven to provide more accurate band gaps and band edge positions than the PBE. For metallic MXenes, we found that their HSE06 band structures are similar to the ones found using the GGA-PBE ~\cite{Heyd-2003, Heyd-2006, Krukau-2006} (Fig. S1 in the
supplementary material). However, for semiconductor MXenes (Sc$_2$COS, Sc$_2$CFN, and Sc$_2$COHN) the band gap obtained when using the hybrid (HSE06) functional increases significantly compared to that obtained by GGA-PBE. The HSE06 functional was also used to calculate the spin-polarized band structure of the MXenes. 

The self-consistent solutions were obtained by employing a ($50\times50\times1$) Monkhorst--Pack~\cite{Mon-76} 
grid of k-points for the integration over the Brillouin zone.
To prevent spurious interaction between isolated layers, a vacuum layer of at least 15 {\AA} was included
along the direction normal to the surface. The elastic tensor 
of each system is derived from the stress–strain approach ~\cite{Le-02}.

\section{Results and Discussions}
Firstly, we fully relaxed the pristine Sc$_2$C. The ground-state structure of pristine Sc$_2$C is found to be a hexagonal crystal structure with equilibrium lattice parameters {\it a}={\it b}=3.29 Å. 
Once the ground-state crystal structure of pristine Sc$_2$C was obtained, we explored different functionalizations of the Sc$_2$C  
MXene monolayer with X= O, F, OH; T= C, N, S. 

%Fig 1
\begin{figure}[h] \centering
	\includegraphics[width=10cm,clip=true]{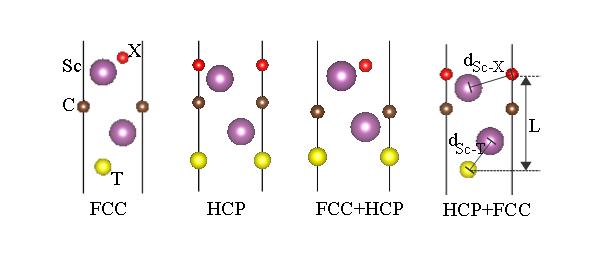}
	\caption{Side views of the four possible models for Sc$_2$CXT (X = O, F, OH; T = C, S, N) MXenes. Sc atoms are shown in purple, C atoms in brown, X (X = O, F, OH) atoms in red, and T (T = C, S, N) atoms in yellow. Please note that stoichiometry is the same for the 4 structures: the seemingly duplicated atoms are an artifact of the periodic boundary conditions in the figure.} 
	\label{models}
\end{figure}

MXT Janus MXenes have four possible functional group models. As presented in Fig. 1, we label them as FCC, HCP, FCC+HCP, HCP+FCC. In the FCC model, each of the X and T atoms is aligned with one of the Sc atoms, whereas in the HCP model these surface atoms are positioned directly below and above the C atoms. As for the FCC+HCP model, the X atoms are placed in the same position as in the FCC model (aligned with the Sc atoms) and the T occupies the same position as in the HCP model (below the C). Reversely, in the HCP+FCC model T atoms are located as in the FCC scheme and the X atoms are placed as in the HCP.  
The calculated lattice parameters ($a$), relative total energies with respect to the most stable model structure for each material ($\Delta$E), 
and formation energies ($\Delta$E$_f$) of these four configuration models are listed in Table 1. Formation energies are calculated to get the most stable configuration based on equation 1: 
\begin{equation}
	\label{eq1}
	\Delta E_{\mathrm{f}}=E_{\mathrm{tot}}(MXT)-E_{\mathrm{tot}}(M)-
	E_{\mathrm{tot}}(X)-E_{\mathrm{tot}}(T)
\end{equation}
\begin{table*}
	\centering 
	\caption{Lattice parameters $a$ (in {\AA}), relative energies with respect to the minimum energy for each model $\Delta$E (in eV), formation energies $E_f$ (in eV), 
		and band gaps (in eV) for the Sc$_2$CXT (X= O, F, OH; T= C, N, S) in the four possible models. 
		Here, M, D, and ID indicate the metallic, direct or indirect bandgap semiconductor character, respectively. $E_f$ and $E_g$ are shown only for the most stable model of each structure.}
	
	\centering \vspace{2mm} \centering
	\footnotesize\setlength{\tabcolsep}{8pt}
	
\begin{tabular}{c| c c| c c| c c| c c| c c }
	\hline
	& \multicolumn{2}{c}{FCC} & \multicolumn{2}{c}{HCP}& \multicolumn{2}{c}{FCC+HCP}& \multicolumn{2}{c}{HCP+FCC} \\ 
	
        	Material & ~$a~$ &  $\Delta$E &~$ a ~$&  $\Delta$E &~$ a~$ &  $\Delta$E &~$ a~$ &  $\Delta$E &$E_f$ & $E_g$ \\
    \hline
	Sc$_2$COC    & 3.33 & 3.15	& 3.54	& 2.66	& 3.58 &	{\bf 0.00} &	3.35 &	3.45 & -4.50 & 	M  \\[0ex]
	 
	Sc$_2$COS  & 3.32&	0.98&	3.51&	0.48&	3.62&	{\bf 0.00}&	3.55&	0.40&	-7.88&	2.45 (ID) \\[0ex]\raisebox{1.5ex}

	%\rowcolor{lightgray}Sc$_2$CON & 3.36	& 2.87&	3.45&	0.55&	3.49&	0.00&	3.51&	2.03&	-8.85&	M  \\[0ex]\raisebox{1.5ex}
%\rowcolor{lightgray}
	Sc$_2$CON & 3.36	& 2.87&	3.45&	0.55&	3.49&	{\bf 0.00}&	3.51&	2.03&	-9.85&	M \\[0ex]
	\hline
	\raisebox{1.5ex}
	
	Sc$_2$CFC  &3.29&	1.83&	3.56&	{\bf 0.00}&	3.62&	0.87& 3.25&	2.39 &		-3.83&	M  \\[0ex]\raisebox{1.5ex}
	
	Sc$_2$CFS  &3.28&	{\bf 0.00}&	3.43&	1.11&	3.25&	0.40&	3.24&	0.64&	-7.77&	M  \\[0ex]\raisebox{1.5ex}
	
	Sc$_2$CFN & 3.36	& 2.87&	3.45&	0.55&	3.49&	{\bf 0.00}&	3.51&	2.03&	-8.95&	1.67 (ID) \\[0ex]
	\hline
	\raisebox{1.5ex}\
	
	Sc$_2$COHC  &3.30	& 1.98&	3.57&	{\bf 0.00}&	3.68&	0.26 & 3.24&	3.86&	-4.49&	M  \\[0ex]\raisebox{1.5ex}

	Sc$_2$COHS  &3.30&	{\bf 0.00}&	3.45&	0.90 & 3.31	& 2.79	&	3.28 & 1.59	&	-4.36&	M  \\[0ex]\raisebox{1.5ex}

	Sc$_2$COHN &3.47&	1.96&	3.48&	{\bf 0.00}&	3.29 &	5.80 & 3.26&	5.29		&-10.43&	1.06 (D) \\
\hline
\end{tabular}
\end{table*}
where E$_{tot}$(MXT), E$_{tot}$(M), E$_{tot}$(X), E$_{tot}$(T)
are the total energy of the fully functionalized MXT (f-MXT), the total 
energy of pristine Sc$_2$C, and the total energy of (1/2) (O$_2$, F$_2$, (OH)$_2$ in the gas phase), C, N, and S (their stable bulk forms), respectively. For the most stable configurations of each material, the thickness of the monolayers (L) and the bond lengths between the surface Sc and the functional group ($d_{Sc-X}$ and $d_{Sc-T}$) have been calculated and are presented in Table 2. From now on we will focus only on the 9 most stable geometries out of the 36 studied potential monolayers.
\begin{figure}[h] \centering
	\includegraphics*[width=9cm,clip=true]{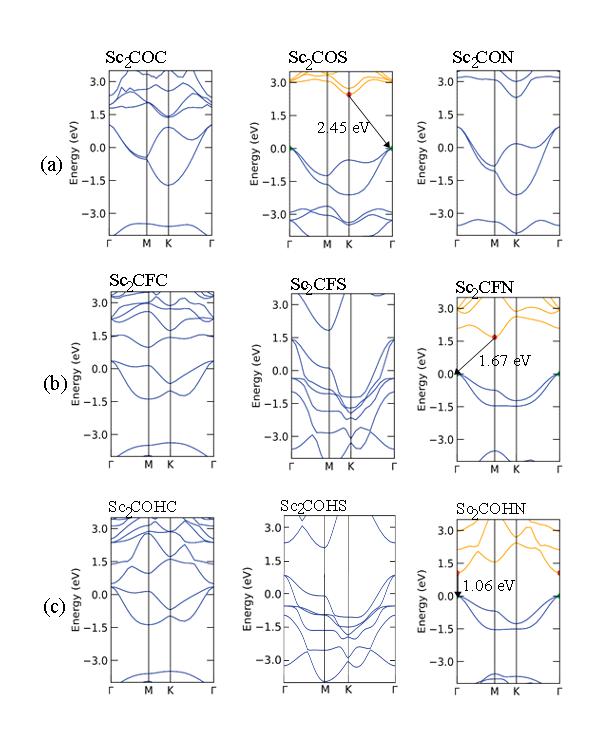}
	\caption{Band structures for the  most stable models of (a) Sc$_2$COT, (b) Sc$_2$CFT and (c) Sc$_2$COHT monolayers, 
		(T= C, S, N). The Fermi energy is set at zero.} 
	\label{models}
\end{figure}
As seen in Table 1, all the formation 
energies (E$_f$) of the MXT under study indicate that the synthesis of these MXenes is 
highly possible (in our convention, more negative formation energy indicates a more 
stable structure.) This stability is further confirmed by the fulfillment of all the materials of the Born criteria for 2D hexagonal materials, as we describe in more detail below. The most favored model is the FCC+HCP, with 4 of the 9 possible structures presenting their most stable geometry in that model. It is followed by the HCP (3 structures) and the FCC (2 structures) models.

\begin{figure}[h] \centering
	\includegraphics*[width=9 cm,clip=true]{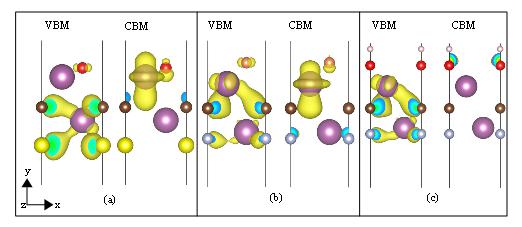}
	\caption{The charge-density isosurfaces of VBM and CBM states of (a) Sc$_2$COS, (b) Sc$_2$CFN, and (c) Sc$_2$COHN. Sc atoms are shown in purple, C atoms in brown, O atoms in red, H atoms in white, F atoms in orange, S atoms in yellow, and N atoms in blue.} 
	\label{models}
\end{figure}

Our results show that, for any given X = O, F, OH, the further functionalization of the opposite surface with nitrogen makes the MXT monolayers more stable than the functionalization with C or S. This 
can be explained by looking at the electronegativity of these species. Indeed, E$_f$ becomes 
more negative with the increase of the electronegativity of T (C, S, or N) 
atoms. The Sc-T bond is formed by the electrons transitioning 
from the Sc to the T atoms, which causes the T atoms to gain 
more electrons, hence further strengthening the Sc-T bond and thus resulting in a  
more negative E$_f$. Due to the stronger bond of Sc-N, the $d_{Sc-N}$ bond length of MXN is also shorter than that obtained for the other functional species $d_{Sc-T}$ (T=C, S), as seen in Table 2. The oxygen functionalization indicates a strong Sc–O interaction, which leads to the shortest $d_{Sc-O}$ bond length.

\begin{table*}
	\centering 
	\caption{Calculated monolayer thickness (L) and bond lengths between the 
    	Sc and the functional groups ($d_{Sc–-X}$, $d_{Sc-–T}$) and between 
    	the Sc and C atom ($d_{Sc-–C}$) of Sc$_2$CXT  (X= O, F, OH; T = C, S, N) MXenes.}
\centering \vspace{2mm} \centering
\begin{tabular}{l|ccccc}
	\hline &&&&& \\[-3ex]
 {Material} & ~$L$({\AA}) ~& ~$d_{Sc–-C}$({\AA})~&~$d_{Sc–-X}$({\AA})~&$d_{Sc–-T}$({\AA})~\\[0.5ex]
	\hline &&&&& \\[-3ex]
	~Sc$_2$COC~ &3.54&	2.24&	2.12&	2.14&	 ~ \\[0.5ex]
	%\hline &&&&& \\[-4ex]
	~Sc$_2$COS~&3.97 & 2.20 & 2.14	& 2.46	&	 ~\\[0.5ex]
	%\hline &&&&& \\[-4ex]
	~Sc$_2$CON &3.66&	2.25 & 2.09	& 2.06	&	~\\[0.5ex]
	\hline &&&&& \\[-3ex]
	~Sc$_2$CFC &3.98&	2.50&	2.35 &	2.08 &	~\\[0.5ex]
	~Sc$_2$CFS & 5.47&	2.32&	2.20&	2.58&	~\\[0.5ex]
	~Sc$_2$CFN &3.84&	2.34&	2.27&	2.03&	 ~\\[0.5ex]
	\hline &&&&& \\[-3ex]
	~Sc$_2$COHC  &  5.00	&2.35&	2.39 & 2.08	&	~\\[0.5ex]
	~Sc$_2$COHS &6.51&	2.33& 2.26 & 	2.58	&	 ~\\[0.5ex]
	~Sc$_2$COHN &5.00&	2.31	&2.36 & 2.04	&	 ~\\[0.5ex]
	\hline
\hline
\end{tabular}
\end{table*}

\begin{figure}[h] \centering
	\includegraphics*[width=9 cm,clip=true]{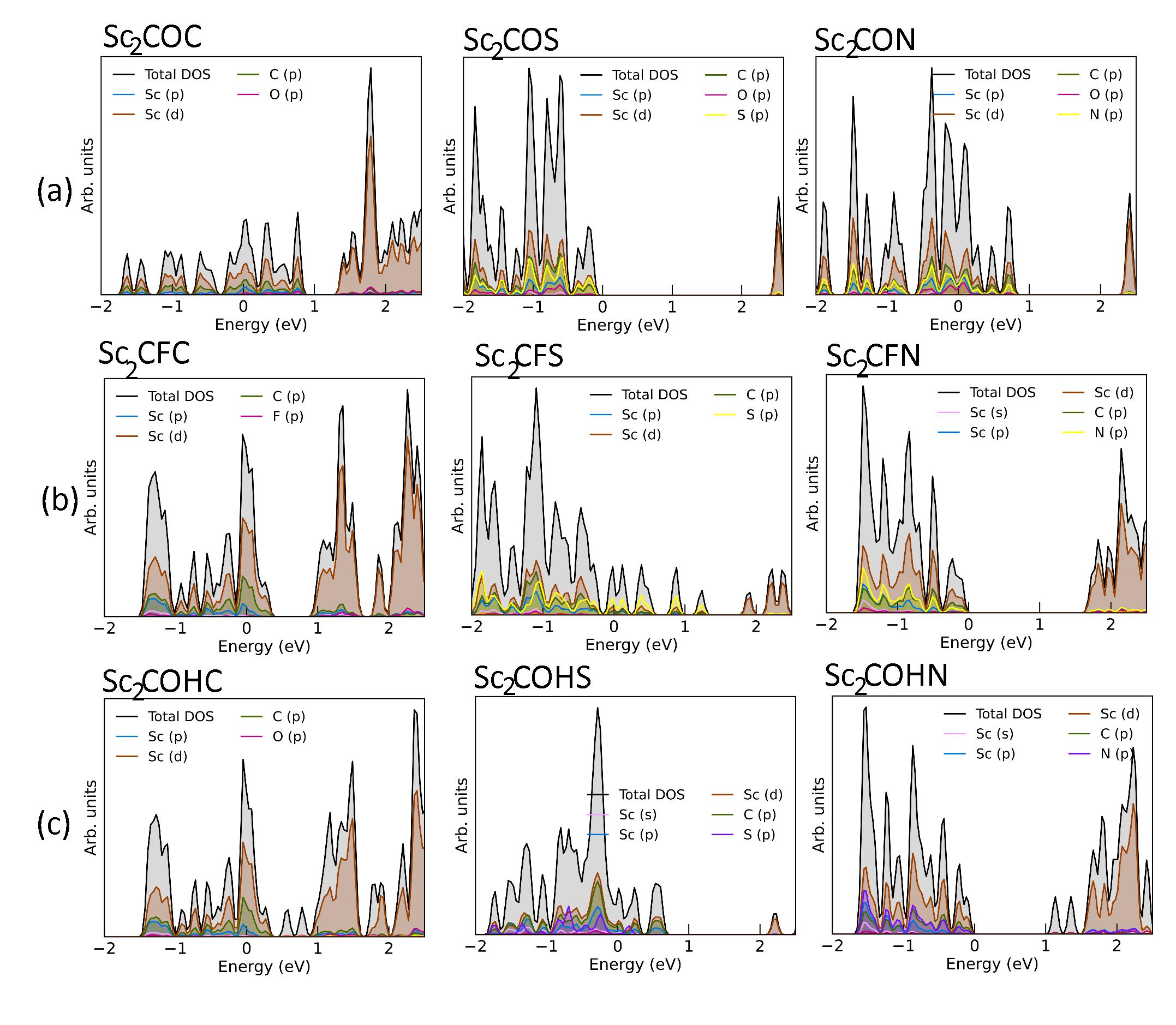}
	\caption{Projected Density of States (PDOS) of the most relevant orbitals for different elements for the (a) Sc$_2$COT, (b) Sc$_2$CFT and (c) Sc$_2$COHT (T=C, S, N) Janus monolayers. The Fermi level is set to zero.} 
	\label{models}
\end{figure}

\begin{figure}[h] \centering
	\includegraphics*[width=10cm,clip=true]{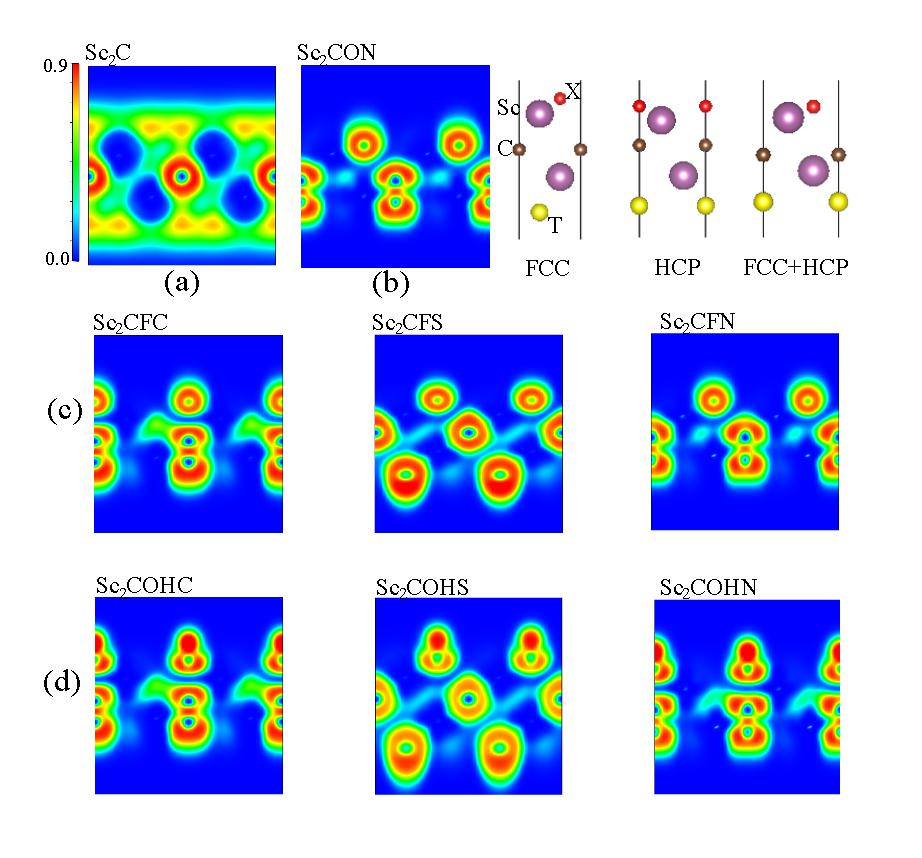}
	\caption{Electron localization function isosurfaces of the (a) Sc$_2$C, (b)Sc$_2$CON, (c) Sc$_2$CFT and (d) Sc$_2$COHT (T = C, S, N) Janus monolayers. The magnitude bar for the ELF is shown in the top left panel. The insert panel at the top right shows the structural models for the Janus monolayers.} 
	\label{models}
\end{figure}
The electronic properties of bare MXenes will be as well affected by surface 
functionalization. While the Sc$_2$C monolayer has a metallic character, previous 
reports ~\cite{Yorulmaz-2016,Wang-2017} show that surface 
functionalization of the Sc$_2$CO$_2$, Sc$_2$CF$_2$ and Sc$_2$C(OH)$_2$ 
monolayers transform them in semiconductors with a band gap energy (E$_g$) of 1.76 eV (indirect band), 
0.94 eV (indirect band), and 0.51 eV (direct band), respectively. To analyze the potential change in the electronic properties of the Janus Sc$_2$CXT studied here, we present in  Fig 2. the corresponding energy band 
structures, where the conduction band is shown in yellow in those cases where a change from the metallic to a semiconductor character is found. Of the 9 Sc$_2$CXT MXenes, six retain their metallic character, while three of them become semiconducting upon functionalization: both Sc$_2$OS and Sc$_2$FN are semiconductors with indirect band gaps of 2.45 and 1.67 eV, 
respectively, while Sc$_2$COHN is a direct band gap semiconductor with a band gap of 1.06 eV. This gap value makes
Sc$_2$COHN a valuable candidate for its use in solar cells, whereas the metallic Sc$_2$CXT monolayers would be much better candidates to act as high-performance electrode materials ~\cite{Lebegue-2009}, 
than, for instance, the semiconducting MoS$_2$ monolayer  ~\cite{Wu-2015, Du-2010}.

To understand the charge distribution of the electronic states near the gap, we have calculated the charge-density isosurfaces of the valence band maximum (VBM) and conduction band maximum (CBM) states of the semiconducting Janus monolayers. As seen in Fig.3, the VBM is primarily derived from the $d$ orbital of the Sc atoms. The bonds between the d orbitals of the Sc atom and the sublayer atoms (C, S, or N) present a $\pi$ bonding character. The CBM is primarily derived from the $p_z$ orbitals of the O atom for Sc$_2$COHN, the Sc $d_z^2$ states and the $p_z$ orbitals of O and F atoms for Sc$_2$COS and Sc$_2$CFN, respectively.

To gain further insight into the electronic
properties we have calculated the partial
density of states (PDOS) for the nine Janus MXenes. As seen from Fig. 4, the valence states are dominated by the Sc (d) states and C (p) states, while the conduction band is dominated by the Sc (d) states for all structures. The strong hybridization between the Sc 3d states and C 2p states is responsible for the mixed
covalent/metallic/ionic character of the Sc–C bonds ~\cite{Sun-04}.

To analyze the electron distributions, we examined the electron localization functions (ELF) ~\cite{Silvi-94} within the (110) plane of the Sc$_2$CXT Mxenes. Since the ELF is similar for both the Sc$_2$COT and Sc$_2$CON MXenes we only show the results obtained from the Sc$_2$CON analysis. From the ELF map in Fig.5 we can see that both the Sc$_2$C and the Sc$_2$CXT surfaces have a cloud of lone-pair electrons. For Sc$_2$CFS and Sc$_2$COHS (the FCC structures) the electrons are mainly located around the F, OH, and S atoms.

In addition to the electronic properties, the mechanical properties of 2D materials play also a critical role regarding their potential technological and practical applications.
We have then investigated the mechanical stability of Janus MXenes according to the Born criteria ~\cite{Born-08}. Based on the Born stability criteria, 2D hexagonal crystals should obey the following conditions: 

\begin{equation}
	\label{eq2}
	\ C_{11}>0, \  C_{66}>0, \ 2*C_{66}=C_{11}-C_{12}, \  and \ C_{11}> |C_{12}| 
\end{equation}

\begin{table*}
	\centering 
	\caption{Elastic constants ($C_{ij}$ in N/m), Young’s moduli (Y in N/m), 
		in-plane stiffnesses (B in N/m), shear moduli (G in N/m), and Poisson’s ratios ($\nu$).}
	
	\centering \vspace{2mm} \centering
	\footnotesize\setlength{\tabcolsep}{8pt}
	
\begin{tabular}{l|ccccccc}
		%\hline
		\hline &&&&& \\[-3ex]
		~Material~ & ~$C_{11}$~ & ~$C_{12}$~&~$C_{66}$~&$Y$~ & ~$B$~ & ~$G$~ & ~$\nu$~\\[0.5ex]
		\hline &&&&& \\[-3ex]
		~Sc$_2$COC~ &~135.180&	45.921	&44.629&119.580&90.590&	44.630&	0.340 ~ \\[0.5ex]
		%\hline &&&&& \\[-4ex]
		~Sc$_2$COS~ & 137.358&	41.169&	48.094	&125.018&	89.300&	48.094&	0.300~\\[0.5ex]
		%\hline &&&&& \\[-4ex]
		~Sc$_2$CON  &157.733&	53.363&	52.185&	139.680&	105.500&	52.185&	0.338~\\[0.5ex]
		\hline &&&&& \\[-3ex]
		~Sc$_2$CFC &117.189	&40.012&	38.588&	103.527&	78.550&	38.588&	0.341~\\[0.5ex]
		~Sc$_2$CFS& 129.088	&36.981&	46.054&	118.494&	82.980&	46.054&	0.286~\\[0.5ex]
		~Sc$_2$CFN &155.300	&38.294&	58.503	&145.857&	96.850&	58.503&	0.247~\\[0.5ex]
		\hline &&&&& \\[-3ex]
		~Sc$_2$COHC& 119.908&	39.593	&40.158&	106.835	&79.730&	40.158&	0.330~\\[0.5ex]
		~Sc$_2$COHS &127.867&	37.777&	45.045	&116.706&	82.770&	45.045&	0.295~\\[0.5ex]
		~Sc$_2$COHN &152.177&	39.448	&56.364&	141.951	&95.780&	56.364&	0.259~\\[0.5ex]
		\hline
\end{tabular}
\end{table*}

The elastic constants $C_{66}, C_{11}, C_{12} $ are tabulated in Table 3. As we can see, 
all Janus MXT MXenes satisfy the mechanical stability criteria, which shows that they are mechanically stable. Due to the hexagonal symmetry, the elastic constant $C_{11}$ is higher than $C_{12}$. Specifically, we have found that the elastic constant $C_{11}$ increases for N-terminated surface MXenes. The $d_{Sc–-N}$ bond lengths of these Sc$_2$CXN (X:O, F, OH) 
MXenes are smaller than the other bond lengths, given in Table 2. The
stronger bonds between Sc and N atoms hence resulting in larger $C_{11}$ elastic constants. 

To further characterize the mechanical properties of these materials, we calculated Young’s modulus $Y$, Poisson’s ratio $\nu$, the  
shear modulus G, and the in-plane stiffness $B$, derived from the elastic constants as ~\cite{Thomas-2018,Wei-2009,Andrew-2012},
\begin{equation}
\label{eq3}
\ Y=(C_{11}-C_{12})/C_{11}, \ \nu =C_{12}/C_{11},
 G=(C_{11}-C_{12})/2, \ B=Y/2*(1-\nu)
\end{equation}

While Poisson’s ratio $\nu$ reflects the mechanical ductility 
and flexibility, Young’s modulus (Y) reflects the stiffness of materials. If Young’s modulus value is high, the material is stiffer. The stiffest
crystal found here is Sc$_2$FN. To put in context these results, we compare our calculated $Y$  with that of other well-known 2D materials. The $Y$ values calculated here for the Janus MXenes family are within the range of 103.527 and 145.857 N/m, much less stiff than that of graphene (341 N/m) ~\cite{Cak-2014} and comparable with the values obtained for the TMDC family ~\cite{Cak-2014}. On the other hand, the $Y$ values of the Janus MXenes studied here are much larger than that of silicene (62 N/m)~\cite{Mortazavi-2017} and germanene (44N/m)~\cite{Mortazavi-2017}. The Poisson’s ratios of the Sc$_2$CXT MXenes studied here are in a range between 0.286 and 0.341 N/m, comparable with those for MoS$_2$ (0.3 N/m) ~\cite{Cak-2014}. This demonstrates that f-MXenes can maintain mechanical flexibility under large strain, making them excellent materials for applications where strain is going to be an issue.

Finally, we have analyzed the magnetic properties of this novel class of Janus MXenes. Since the recent discovery of magnetic order in 2D materials \cite{Gong-2017,Huang-2017,Wang-2016}, the possibility of exploring new physical phenomena and designing truly novel 2D devices has become a reality that is already being explored experimentally \cite{Zhong-2017}. In particular, the search for half-metallic magnets (HMMs) and spin-gapless semiconductors (SGSs) \cite{Wang2008} has been very intense, due to their intriguing tunable physical properties, and we have explored this possibility in this work. It is worth mentioning that the Sc$_2$CO$_2$ monolayer actually exhibits ferroelectric polarization \cite{Chandrasekaran2017}, and the exploration of the ferroelectric properties of the  Janus  Sc$_2$CXT (X = O, F, OH; T = C, S, N) family needs to be addressed in the near future. 

To explore the magnetic character of the Sc$_2$CXT (X = O, F, OH; T = C, S, N), we have calculated the total energies of the ferromagnetic (FM), non-magnetic (NM), and antiferromagnetic (AFM1 , AFM2, and AFM3) spin ordering of the nine most stable materials studied in this work in a 2$x$2 supercell. This is the minimum supercell size needed to properly study all the possible AFM configurations, as depicted in Figure 6. 

Our results are tabulated in Table 4 for the magnetic structures and Table S1 for the NM ones. The ground state of all structures is NM except for Sc$_2$CFC and Sc$_2$COHC. For Sc$_2$CON, the total energy difference between the FM and NM state is very small, which means that the ground state could easily be unstable in a specific environment. Then, the transition to an FM ground state might occur under certain conditions, and for this reason we have also explored the magnetic properties of Sc$_2$CON (presented in SI). Our results show that Sc$_2$CFC is FM with a supercell magnetic moment of 3.99 $\mu$B, while the ground state of Sc$_2$COHC is AFM1 with a supercell magnetic moment of 3.98 $\mu$B. If the ground state of Sc$_2$CON became magnetic, it would be FM with a supercell magnetic moment of 2.24 $\mu$B. The magnetic moments of the individual atoms for Sc$_2$CFC and Sc$_2$COHC are also given in Fig S2.

\begin{table*}
	\centering 
	\caption{The total energies (in eV) of non-magnetic E$_{NM}$, ferromagnetic E$_{FM}$ and antimagnetic E$_{AFM1}$, E$_{AFM2}$ and E$_{AFM3}$  states (eV), Curie or Néel temperatures (K), and Magnetic Anisotropy Energy (MAE) ($\mu$ eV per Sc atom ) for the high symmetry directions of Sc$_2$CFC and Sc$_2$COHC for a
	2x2x1 supercell.}
\centering \vspace{2mm} \centering
\begin{tabular}{l|ccccccccccc}

	\hline &&&&&&&&&&& \\%[-2ex]
    ~Material~& E$_{FM}$ &E$_{NM}$
    & E$_{AFM1}$  
    & E$_{AFM2}$ &E$_{AFM3}$
    &T$_{C/N}$ & (100)& (010)& (110)& (001)& (111) \\[0.5ex]
	\hline &&&&&&&&&&& \\%[-3ex]
%	~Sc$_2$CON &-166.402 & -166.406 &	-166.398 &		-166.398 & 	-166.397 & 31 & 0 & 0 & 0 & 3 & 1 ~\\[0.5ex]
	
	~Sc$_2$CFC~ &-147.578&	-147.435& 	-147.576& 	-147.576& 	-147.576&	22 & 32 & 	32 & 32 & 24 & 0	~\\[0.5ex]

	~Sc$_2$COHC  & -171.452 & -171.284	&	-171.453  &	-171.452  & -171.453 &	9 & 8 & 8 & 8 & 0 & 5 ~\\[0.5ex]

	\hline
\hline
\end{tabular}
\end{table*}

\begin{figure}[h] \centering
	\includegraphics*[width=9 cm,clip=true]{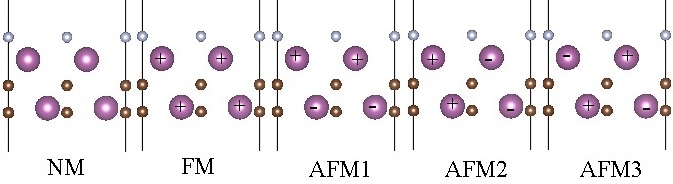}
	\caption{The non-magnetic and four different magnetic configuration models: NM, FM, AFM1, AFM2, AFM3 considered in the calculation.
	Magnetic state configurations. In Sc atoms, “$+$” represents spin-up, and “$-$“ represents spin-down.} 
	\label{models}
\end{figure}
\begin{figure}[h] \centering
	\includegraphics*[width=8cm,clip=true]{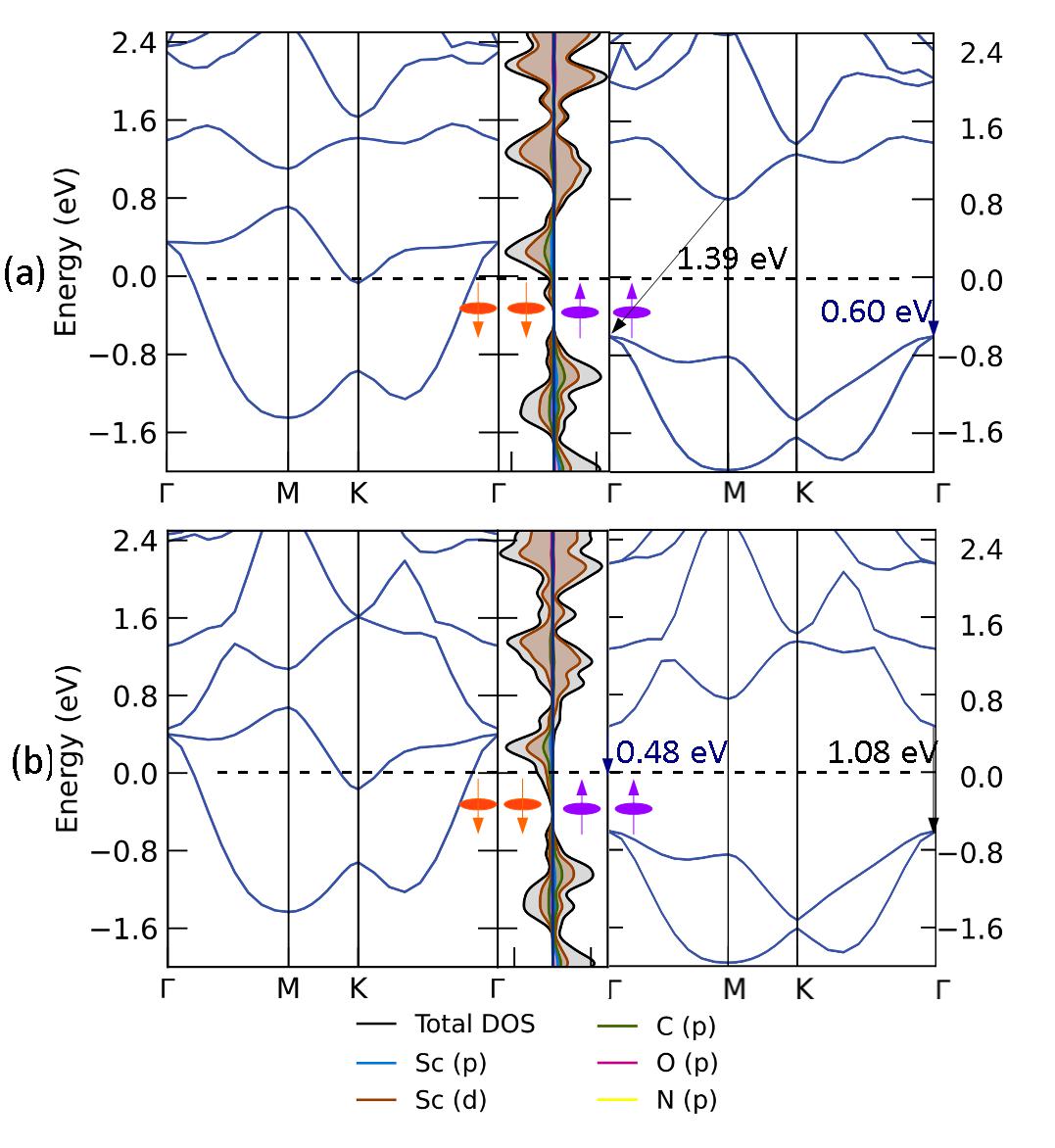}
	\caption{The band and spin-resolved density of states structures of (a)Sc$_2$CFC, and (b)Sc$_2$COHC. The Fermi level is set at 0 eV. Bands and DOS for spin-down (up) are shown on the left (right) panels. The black and blue arrows show the HSE06 band gap and the fundamental band gap, respectively.} 
	\label{models}
\end{figure}

The spin-dependent electronic band structure and DOS for the ground states of Sc$_2$CFC (FM) and Sc$_2$COHC (AFM) are shown in Fig. 7 ( the spin-polarized bandstructure and DOS of Sc$_2$CON in the FM state are shown in Fig. S3.). Remarkably, their band structures present a half-metallic character, where the majority (spin-up) electrons show semiconducting behavior with an HSE06 half-metallic (or fundamental) energy gap of 0.60 eV and 0.48 eV, respectively, while the minority (spin-down) electrons present a metallic behavior. Half-metallicity is a very much searched for quality in nanomaterials, since, provided the half-metallic energy gap of the majority electrons is wide enough, the electronic transport in this material would be totally spin-polarized, dominated by the minority-spin electrons. Similar results have been found for other Janus materials \cite{Zhang2021,He2019,He2016,Akgenc2020, Zhong-20, Wang-19, Zuntu-21, Wang-16}. In our case, the values of the fundamental bandgaps found for Sc$_2$CFC and Sc$_2$COHC are larger than those typically found for chalcogenides and some types of perovskites, making them much more suitable for spintronics applications. A small fundamental band gap means that the half-metallic character could disappear under certain conditions, such as strain, so having a half-metallic gap larger than at least half an eV is essential to preserve this important property.

The Curie or Néel temperature (T$_C$ or T$_N$) of Sc$_2$CFC and Sc$_2$COHC was calculated based on the mean-field theory approximation (MFA), which is estimated by k$_B$T$_{(MFA)}$= (2/3) $\Delta({E})$ ~\cite{Jin-09}. Here, k$_B$ and $\Delta({E})$ are, respectively, the Boltzmann constant and, for Sc$_2$CFC (which has a FM ground state), the energy difference between FM and AFM (all the AFM energies are equal), or, for Sc$_2$COHC (which has a AFM ground state), the energy difference between FM and AFM1 (the energy of AFM1 is the same as the AFM3, and smaller than AFM2). The T$_C$ of Sc$_2$CFC, and T$_N$ of Sc$_2$COHC  were calculated to be 22 K, and 9 K, respectively. Although these values are low, the energy difference between the magnetic and non-magnetic states in both systems is large enough to likely maintain the magnetic features at a temperature close to room temperature, as in the case of  
%(-0.14 eV and -0.17 eV, for Sc$_2$CFC and Sc$_2$COHC, respectively) 
Sc$_2$NO$_2$ ~\cite{Liu-17, Khazaei-13}.

The Magnetic Anisotropy Energy (MAE) determines the FM ordering in 2D materials. 
We have examined the impact of 
spin-orbit 
coupling 
(SOC) 
and electron 
localization on the magnetic properties of Sc$_2$CFC, and Sc$_2$COHC. The MAE is defined as the difference in energy between the system with a given spin orientation, and the energy of the most stable spin orientation, which is called an easy axis. To understand the nature of FM ordering in these materials, we consider the in-plane a (100), b (010), and a+b (110), and the out-of-plane c (001) and (111) high-symmetry axes. 
Since the changes in energies are very low, we used a dense k-mesh of 21 × 21 × 1 for more accurate energy convergence. As we can see in Table 4, both Sc$_2$CFC and Sc$_2$COHC exhibit an out-of-plane easy axis, which are the (111) and (001), respectively. The calculated MAEs are 32 and 8 $\mu$eV (per Sc atom) for (100), (010), and (110) directions. These values are similar to previous calculated results for 2D
materials ~\cite{Hu-16}, but larger than those found for Fe (1.4 $\mu$eV per Fe atom) and Ni (2.7 $\mu$eV per Ni atom)
bulks ~\cite{Daalderop-90}.   

These features make Sc$_2$COHC, and especially Sc$_2$CFC, very promising candidates for spintronic applications.

\section{Summary}

We have presented a full
{\it ab initio} study of the atomic geometry, stability, electronic and magnetic properties of a new family of functionalized Janus MXenes, Sc$_2$CXT (X:O, F, OH; T:C, N, S). Following energetic and mechanical criteria, we have shown the high stability of these materials. In some cases, functionalization with the appropriate groups changes the metallic character of the original Sc$_2$C MXene into a semiconducting one, with either a direct or indirect band gap, depending on the actual surface termination. In addition, the analysis of the mechanical properties proves the excellent mechanical flexibility of f-MXenes under large
strain. Finally, we found that three of the Janus MXenes under study in this work present ferro- or antiferromagnetic characteristics, particularly a half-metallic character with a significant fundamental energy gap. 
All these remarkable properties make the Sc$_2$CXT (X:O, F, OH; T:C, N, S) Janus MXene family a valuable candidate for future optoelectronic and spintronic nanodevices.

\section*{Author Contributions}
Sibel Özcan: Conceptualization (lead); formal analysis (lead); writing – original draft (lead); writing – review and editing (equal). Blanca Biel: Conceptualization (supporting); writing – review and editing (equal); funding acquisition (lead).

\section*{Conflicts of interest}
There are no conflicts to declare.

\section*{Acknowledgments}

S.Ö. and B.B. kindly acknowledge financial support by the Junta de Andalucía under the Programa Operativo FEDER P18-FR-4834. B.B. also acknowledges financial support from AEI under project PID2021-125604NB-I00. The Albaicín supercomputer of the University of Granada and TUBITAK ULAKBIM, High Performance and Grid Computing Center (TRUBA resources) are also acknowledged for providing computational time and facilities.

%%%END OF MAIN TEXT%%%


\end{document}